\documentclass[aps,pra,groupedaddress,tightenlines, twocolumn]{revtex4}
\usepackage{amsfonts, bbm}
\usepackage[english]{babel}
\usepackage[mathscr]{eucal}

\usepackage{epsfig}

\def\tr{{\rm Tr}}

\def\id{{I}}
\def\piv{\mbox{\boldmath$\pi$}}
\def\eqref#1{~(\ref{#1})}

\begin{document}

\title{Optimal signal states for quantum detectors}

\author{Ognyan Oreshkov$^{1,2}$, John Calsamiglia$^{1}$, Ramon Mu\~{n}oz-Tapia$^{1}$, and Emili Bagan$^{1,3,4}$}

\address{${^1}$F\'isica Te\`orica: Informaci\'o i Fen\`omens Qu\`antics, Universitat
Aut\`{o}noma de Barcelona, 08193 Bellaterra (Barcelona), Spain\\
${^2}$QuIC, Ecole Polytechnique, CP 165, Universit\'{e} Libre de Bruxelles, 1050 Brussels, Belgium\\
${^3}$Department of Physics, Hunter College of the City University of New York, 695 Park Avenue, New York, NY 10021, USA\\
${^4}$HET Group, Physics Department, Brookhaven National Laboratory, Upton, NY 11973, USA\\}
\begin{abstract}

Quantum detectors provide information about the microscopic properties of quantum systems by establishing correlations between those properties and a set of macroscopically distinct events that we observe. The question of how much information a quantum detector can extract from a system is therefore of fundamental significance. In the present paper we address this question within a precise framework: given a measurement apparatus implementing a specific POVM measurement, what is the optimal performance achievable with it for a specific information readout task, and what is the optimal way to encode information in the quantum system in order to achieve this performance? We consider some of the most common information transmission tasks---the Bayes cost problem, unambiguous message discrimination, and the maximal mutual information. We provide general solutions to the Bayesian and unambiguous discrimination problems. We also show that the maximal mutual information is equal to the classical capacity of the quantum-to-classical channel describing the measurement, and study its properties in certain special cases. For a group covariant measurement, we show that the problem is equivalent to the problem of accessible information of a group covariant ensemble of states. We give analytical proofs of optimality in some relevant cases.
The framework presented here provides a natural way to characterize generalized quantum measurements in terms of their information readout capabilities.

\end{abstract}

\maketitle



\section{Introduction}


Quantum detectors provide the interface between the microscopic world of quantum phenomena and the world of macroscopically distinct events that we observe. A quantum detector is a device that interacts with the system under observation in a way that establishes correlations between certain properties of the system and a set of macroscopically distinct (orthogonal) states of the device. A general quantum detector can be described by a positive operator-valued measure (POVM), i.e., a set of positive operators $\{E_i\}$, $E_i\geq 0$, $i=1,..., M$, summing up to the identity, $\sum_iE_i=I$. For an input state $\rho$, the probability that the measurement yields outcome~$j$ is
given by the Born rule, $p_j(\rho)=\tr\{\rho E_j\}$.

A natural question is to what extent a given quantum detector is able to provide information about the system it is used to observe. This question can be conveniently formulated in the context of a quantum communication scenario, where a sender (Alice) tries to send messages to a receiver (Bob) who is constrained to read those messages using the quantum detector in question. Concretely, let the source of classical information that Alice wants to
communicate to Bob be characterized by a probability
distribution $\pi_i>0$, $i=1,...,N$, $\sum_i\pi_i=1$, that specifies
the probability of each classical message $i$. Alice encodes
the different messages into quantum states via an encoding map
$i\rightarrow \rho_i$, and Bob reads the information by
performing the POVM measurement. If there are no constraints on the way Alice can prepare the signal states and these states can reach Bob undisturbed (i.e., Alice and Bob are connected through a noiseless channel), then the optimal performance they can achieve for a given task can be regarded as quantifying the readout capabilities of the measurement with respect to that task. In this respect, a problem of primary importance is to find the optimal encoding (or signal states $\rho_i$) for which the detector achieves its optimal performance.

The problem just outlined bears strong similarities to the problem of quantum state discrimination \cite{Helstrom74, Davies78, Ivanovic87, Dieks88, Peres88, Fuchs96, Chefles98, Raynal05}, where the encoding of Alice is fixed and Bob's task is to decide which message he has received by optimizing his measurement. In fact, we will see below that the two problems can be regarded as dual to each other due to the symmetry that exists between the input ensembles and the POVM measurements. This allows one to adopt results from quantum state discrimination to the problem at hand. However, since in quantum state discrimination the space over which we optimize is more constrained due to the completeness relation $\sum_iE_i=I$, it turns out that in many cases the problem of optimal signal states for quantum detectors is easier to solve.

Besides its application for characterizing detectors, the problem considered here is of natural practical interest for quantum communication, since generating different signal states \cite{Lundeen} can be experimentally more accessible than performing different measurements. A quantum detector is usually fixed, while a preparation device, although possibly also based on a fixed (but nondestructive!) measurement, can be used together with post-selection, which provides additional flexibility to the preparation process. Furthermore, in the case of communication through a noiseless channel, any operation at the receiver's side prior to the detector can be equally done as part of the preparation strategy.

In this paper, we consider the above problem from the perspective of three different information transmission tasks---the task of optimal Bayes cost message discrimination (of which the well known problem of minimum error discrimination is a special case), unambiguous message discrimination, and the maximal mutual information. Due to the simplification mentioned above, we are able to provide solutions to the Bayesian and unambiguous discrimination problems in the general case. For the maximal mutual information, we show that this quantity is equal to the classical capacity of the quantum-to-classical channel corresponding to the measurement, which we term ``\textit{capacity of the measurement}''. This quantity provides a general figure of merit for the information readout capabilities of a detector. Based on its relation to the accessible information \cite{Fuchs96}, we prove a result similar to Davies's theorem \cite{Davies78} (Proposition 2), which shows that the optimal ensemble can be chosen to consist of $d^2$ pure states, where $d$ is the dimension of the system. For a group covariant measurement, we obtain that the problem is equivalent to that of accessible information of a group covariant ensemble of states. We apply our results to the case of a noisy two-level symmetric informationally complete measurement, for whose capacity we give analytical proofs of optimality.

\section{The Bayes cost problem \label{sec:bayes}}

In the Bayes cost problem, one is interested in minimizing an average cost function of the form
\begin{equation}
C(P)=\sum_{ij}C_{ij}P_{ij},\label{cost}
\end{equation}
where $P_{ij}=\tr(\pi_i\rho_iE_j)$ are the joint probabilities for
input $i$ and measurement outcome $j$, and $C_{ij}\geq 0$ are the
elements of the cost matrix ($C_{ij}$ is the cost of choosing
hypothesis $j$ when hypothesis $i$ is true). In what we will refer
to as the \textit{straight} version of this problem, one assumes
that the encoding $i\rightarrow \rho_i$ is given, and the task is to
find the measurement $\{E_j\}$ that minimizes the quantity in
Eq.\eqref{cost} \cite{Helstrom74}. An example of a Bayes cost
problem is that of \textit{minimum error discrimination}, i.e.,
minimizing the probability for incorrectly identifying the message.
In this case, the probability for an error is given by
$p_{\textrm{err}}=\sum_{i\neq j}P_{ij}$, i.e., the elements of the
cost matrix are $C_{ij}=1-\delta_{ij}$.

Here we are concerned with the opposite scenario which we will refer
to as the \textit{reverse} problem: we assume that the receiver
has an apparatus that implements a particular POVM measurement, and
we ask what the optimal way to encode the classical messages
into quantum states is so that, using only the given POVM
measurement and possibly some side information processing, the
receiver will identify the message at the lowest cost.
This side information processing involves finding the optimal way of choosing hypothesis $i$ when the measurement outcome $k$ takes place, and includes the possibility of following a mixed strategy, i.e., assigning a hypothesis $i$ randomly according to some prescribed probability distribution, which might of course depend on the outcome $k$.
 In other words, the receiver can use the
given POVM $\{E_k\}$ to obtain new POVM measurements with elements
of the form
\begin{equation}
\widetilde{E}_j=\sum_k p(j|k)E_k,\qquad \sum_j p(j|k)=1\;\mbox{\rm for all $k$},
\label{ebc31.03.11-1}
\end{equation}
where $0\leq
p(j|k)\leq 1$ are conditional probabilities.

Up to renormalization of the cost matrix, we can assume that $0\leq C_{ij}\leq 1$. Hence, the problem is equivalent to that of maximizing the quantity
\begin{equation}
B(P)\equiv 1-C(P)=\sum_{ij}(1-C_{ij})P_{ij}\equiv \sum_{ij}B_{ij}P_{ij}, \label{cost2}
\end{equation}
where
\begin{equation}
0\leq B_{ij}\leq 1, \hspace{0.2cm} \forall i,j.
\end{equation}

For a given POVM measurement $\{E_i\}$, consider some encoding and decoding strategies given by the map $i\rightarrow\rho_i$ and the conditional probability distribution $ p(j|k)$, respectively. For these strategies, the quantity $B(P)$ reads
\begin{equation}
B(P)=\sum_{ijk}B_{ij}\pi_i p(j|k)\tr(\rho_iE_k).
\end{equation}
Define $j^{\ast}(k)$ to be a value of $j$ for which the quantity $\sum_iB_{ij}\pi_i\tr(\rho_iE_k)$, for a fixed $k$, is maximal (if there are two or more such values, we can pick any one of them). Then,
\begin{equation}
B(P)\leq \sum_{ik}B_{ij^{\ast}(k)}\pi_i\tr(\rho_iE_k),
\end{equation}
which is achievable by choosing $ p(j|k)=\delta_{jj^{\ast}(k)}$.

We see that for the purpose of achieving the maximum in Eq.~\eqref{cost2}, the receiver does not need a mixed strategy, i.e., the maximum can be achieved by choosing all conditional probabilities $ p(j|k)$ to be either $0$ or $1$. This means that the receiver can associate more than one measurement outcome $E_k$ with the same hypothesis $j$, but it does not help to associate two or more hypotheses with the same outcome. Note that this means, in particular, that in the case when the number of possible messages $N$ is greater than the number $M$ of different outcomes of the POVM, the best strategy is not to attempt to detect certain messages at all. In fact (see below), even when $M\leq N$, it may be advantageous to group different POVM elements for the detection of a single state.

Let $K_j$ denote the set of those indices $k$ for which~$j^{\ast}(k)=j$, i.e., the indices $k$ for which the outcomes~$E_k$ are associated with hypothesis $j$. Note that the sets $K_j$ are non-intersecting as shown above and that some sets may be empty. In other words, the set of possible assignments corresponds to that of all possible ways to distribute $M$ elements into $N$ groups $\{ K_{j}^\alpha\}_{j=1}^N$, where the index $\alpha$ labels each of the $N^M$ distributions.
Then for any such choice we have
\begin{equation}
B_{\alpha}(P)=\max_{\{\rho_i\}}\sum_i\pi_i\tr (\rho_i\sum_jB_{ij}\tilde E^\alpha_{j}),
\end{equation}
where $\tilde E^\alpha_{j} =\sum_{k\in K^\alpha_j}E_k$ .
The maximum of this quantity is achieved when each of the signal
states $\rho_i$ is chosen to be an eigenstate corresponding to the
maximal eigenvalue of the operator $\sum_jB_{ij}\widetilde E^\alpha_{j}$,
which we will denote by $\lambda^{\textrm{max}}(\sum_jB_{ij}\widetilde E^\alpha_{j})$. Hence, we
can write
\begin{equation}
B(P)=\max_{\alpha} \piv \cdot \mathbf{s}_{\alpha},\label{Blambdamax}
\end{equation}
where we have defined the vectors $\piv=\{\pi_{1},\ldots,\pi_{N}\}$ and
 $$\mathbf{s}_{\alpha}=\{\lambda^{\textrm{max}}(\sum_jB_{1j}\widetilde E^\alpha_{j}),\ldots,\lambda^{\textrm{max}}(\sum_jB_{Nj}\widetilde E^\alpha_{j}\}.$$
We thus see that the problem reduces to that of finding the sets $K_j$ for which the quantity in Eq.~\eqref{Blambdamax} is maximal.  The corresponding partition specifies which outcomes $k$ of the POVM measurement the receiver has to associate with a given classical message $j$. The optimal encoding strategy is to encode each classical message $i$ into an eigenstate $\rho^{\textrm{max}}_i$ corresponding to the maximal eigenvalue of $\sum_jB_{ij}\widetilde E_j^\alpha$ (note that these states can always be chosen to be pure).

In general, the optimal grouping  $\alpha^*$ of POVM elements,  
$\alpha^*=\arg\max_\alpha \piv \cdot \mathbf{s}_{\alpha}$,
will depend on the given priors $\piv$. The region in the corresponding simplex where one particular grouping is optimal defines a polytope,
or, more precisely, a convex polytope when restricted to the region $\pi_{1}\geq \pi_{2}\geq\ldots\geq \pi_{N}$ (throughout the paper this ordering of prior probabilities will be always assumed), i.e., if $\piv \cdot (\mathbf{s}_{\alpha^*}-\mathbf{s}_{\alpha})\geq 0$ and $\piv'\cdot (\mathbf{s}_{\alpha^*}-\mathbf{s}_{\alpha})\geq 0$, then for $0\le p\le1$ one has  $\left[p\piv+(1-p)\piv'\right]\cdot (\mathbf{s}_{\alpha^*}-\mathbf{s}_{\alpha})\geq 0$.

The described optimization procedure involves calculating and comparing a finite set of quantities. In contrast, the straight version of the problem in the general case is a linear program that requires maximization over a continuous set. Even though the task of finding the optimal encoding for a given decoding POVM exhibits an apparent similarity with the problem of finding the optimal POVM for a given encoding (see the symmetry of the cost function \eqref{cost} with respect to interchanging the POVM elements and the input states), an important difference between the straight and reverse problems is that the quantities over which we maximize in the straight version have to satisfy the constraint $\sum_jE_j=I$, whereas in the reverse case there is no constraint on the signal states $\rho_i$.

Observe that in the case when $N<M$, the above optimal strategy
requires at least one of the messages to be associated with multiple
measurement outcomes. However, as mentioned earlier, even in the
case when $N\geq M$, it may be advantageous to associate more than
one outcome of the POVM with the same state. For example, in the
problem of minimum error discrimination, two POVM elements may have
very similar (or even identical) maximal eigenvalues and
corresponding maximal eigenstates, but all prior probabilities of
the different input messages may differ significantly. Then it is
not difficult to see (see examples in the last section)  that associating the two measurement outcomes
in question with two different messages would be worse than
associating both of them with one of the messages---the one that has
a higher prior probability.

Note that the special case of minimum error discrimination with a given POVM has been previously studied in Ref.~\cite{Elron} as part of the problem of optimal encoding of classical information in a quantum system for minimal error discrimination when both the encoding and the measurement can be optimized. However, the solution provided in Ref.~\cite{Elron} for a fixed POVM is not truly optimal since it assumes that different outcomes must be associated with different states.

We remark that in certain cases it may be possible to simplify the
general procedure described above. For example, in the problem of
minimum error discrimination, when the prior distribution is flat,
$\pi_i=1/N$, $i=1,...,N$, and $M\leq N$, all we need to do is encode
$M$ of the $N$ different possible messages into the eigenstates
corresponding to the maximal eigenvalues of the different POVM
elements. In this case, associating multiple measurement outcomes
with the same message does not help since
$(1/N)\lambda^{\textrm{max}}(E_j+E_k)\leq
(1/N)\lambda^{\textrm{max}}(E_j)+(1/N)\lambda^{\textrm{max}}(E_k)$.

For a binary source, the minimum error probability can be written in a particularly simple form.
In this case, the POVM grouping is $\{\widetilde{E}^\alpha,I-\widetilde{E}^\alpha\}$.
We start discussing the unbiased
case  (i.e., $\pi_{1}=\pi_{2}=1/2$)  for which
\begin{eqnarray}
p^\alpha_s=\max_{\{\rho_1,\rho_2\}}\frac{1}{2}[\tr \widetilde{E}^\alpha \rho_1 +\tr(I- \widetilde{E}^\alpha) \rho_2]\nonumber\\
               =   \frac{1}{2}[1+ \max_{\{\rho_1,\rho_2\}}\tr \widetilde{E}^\alpha(\rho_1-\rho_2)].
\end{eqnarray}

The maximum occurs when $\rho_1$ and $\rho_2$ are the states corresponding to the largest and lowest eigenvalue of $\widetilde{E}^\alpha$, respectively. The difference between these two values is known as the spread of a matrix, defined for a generic matrix $A$ as  $\mathrm{Spr}(A)=\max_{ij}|\lambda_i-\lambda_j|$, where $\lambda_i$ are the characteristic roots of $A$ \cite{marcus}. Hence,
\begin{equation}
p_s=\frac{1}{2}\left[ 1+\max_\alpha \mathrm{Spr}(\widetilde{E}^\alpha) \right].
\end{equation}
Notice the resemblance with the well known Helstrom's state discrimination formula~\cite {Helstrom74}, where the trace-distance has been replaced by the (semi-norm) spread.

From Eq.~\eqref{Blambdamax}, the success probability for arbitrary priors reads
\begin{equation}
p_s=\max_\alpha \{\pi_1 \lambda^{\rm max}(\widetilde E^\alpha)+\pi_2 [1-\lambda^{\rm min}(\widetilde E^\alpha)]\}.
\end{equation}
%
%
%
It is clear that when one signal is given with a prior probability larger than the success probability attained by a  two-outcome POVM $\{E,\id-E\}$,  it pays to assign all outcomes to the most probable signal. In other words, the measurement does not add  information to our prior knowledge, and the optimal grouping results in the trivial POVM~$\{\id,0\}$.
The transition 
occurs at $\pi_1=p_s$. More explicitly, the trivial POVM is optimal if
\begin{equation}
\pi_1\geq \frac{1-\lambda^{\rm min}(E)}{[1-\lambda^{\rm min}(E)]+[1-\lambda^{\rm max}(E)]}.
\end{equation}
Notice that if $\lambda^{\rm max}(E)=1$, it is always advantageous to perform the measurement, irrespectively of the prior probabilities.

\section{Unambiguous message discrimination}

Unambiguous quantum state discrimination
\cite{Ivanovic87,Dieks88,Peres88,Chefles98, Raynal05} concerns the task of
identifying which out of a set of possible states one has received in such a way as to ensure no error whenever a conclusive answer is given. In general, such conclusive answers cannot always be given, and the problem consists in maximizing the probability with which they occur.

Let $\{E_i\}$ be the POVM the receiver has been provided with and let us allow, as in the previous section, some side information processing that will result in  new POVMs,~$\widetilde{E}_j$ 
[see Eq.~(\ref{ebc31.03.11-1})].
For the
purpose of unambiguously identifying a given set of messages
$i=1,...,N$, encoded in the quantum states $\rho_i$,
$i=1,...,N$, these POVMs must consist of $N+1$ elements:
$\widetilde{E}_1,...,\widetilde{E}_N$, representing the conclusive answers, and an additional element $\widetilde{E}_?$ that represents the inconclusive one.
It must hold that
\begin{equation}
\tr(\widetilde{E}_i\rho_j)= \lambda_j\delta_{ij}, \hspace{0.2cm}i,j=1,...,N,\label{condun}
\end{equation}
since errors are not allowed in conclusive answers.
Any of the elements $\widetilde{E}_1,...,\widetilde{E}_N,\widetilde{E}_?$ can be the zero operator as a special case.

One can readily see that all the conditional probabilities~$p(j|k)$ that define $\{\widetilde{E}_i\}$ in terms of the original POVM through Eq.~(\ref{ebc31.03.11-1})
can be taken to be either~$0$ or~$1$, as for the Bayes cost problem, i.e., $\{\widetilde{E}_i\}$ can be taken to be sums of certain subsets
of the original POVM elements. 
This is so because there is no way one can unambiguously  identify two or more messages that have been associated with a given $E_i$
if the corresponding outcome takes place.
(If some outcome $i$ occurs  with zero probability, we can add $E_i$ to any of the elements
$\widetilde{E}_1,...,\widetilde{E}_N,\widetilde{E}_?$, as this
would not change the probabilities of the respective outcomes.)
Similarly, if $E_k$ is randomly associated  with both a given
message $i$ [i.e., $0<p(i|k)$] and the inconclusive answer
[i.e., $0<p(?|k)$%
], the probability of success would increase with the choice $p(i|k)=1$.

Thus, for the
unambiguous discrimination of $N$ input states $\rho_i$,
$i=1,...,N$, each occurring with prior probability $\pi_i$, consider some grouping of the original POVM elements into $N+1$ elements,
$\widetilde{E}_1^\alpha,...,\widetilde{E}_N^\alpha, \widetilde{E}_?^\alpha$,
where, as in the previous section, $\alpha$ labels the various grouping possibilities. Condition \eqref{condun} requires that
$\rho_i\in {\mathscr K}_i^\alpha\equiv \cap_{j\not=i}^{N}\ker \widetilde{E}_j^\alpha$ for
each~$i$.
Conversely, if each $\rho_i$ is chosen
to belong to this intersection (assuming it is non-empty), then unambiguous discrimination would be possible with probability
\begin{equation}
p_{s}^\alpha=\sum_{i=1}^{N}\pi_i\tr(\widetilde{E}_i^\alpha\rho_i)\label{P}.
\end{equation}
Let $P_i^\alpha$ denote the projector on  ${\mathscr K}_i^\alpha$. Note that this projector can be easily computed because  ${\widetilde E}_j^\alpha\ge0$ implies that ${\mathscr K}_i^\alpha=\ker(\sum_{j\not=i}^N\widetilde{E}_j^\alpha)$. Since $\rho_i=P_i^\alpha\rho_i P_i^\alpha$, Eq.~(\ref{P}) can be written as
\begin{equation}
p_{s}^\alpha=
\sum_{i=1}^{N}\pi_i\tr\left[(P_i^\alpha\widetilde{E}_i^\alpha P_i^\alpha)\rho_i\right],
\end{equation}
and we can maximize each of the traces
by choosing~$\rho_i$ to be an eigenstate of $P_i^\alpha\widetilde{E}_i^\alpha P_i^\alpha$
with maximal eigenvalue. Let us
denote this eigenvalue by $\lambda^{\textrm{max}}(P_i^\alpha\widetilde{E}_i^\alpha P_i^\alpha)$.
Then, we have
\begin{equation}
p_{s}^\alpha=\sum_{i=1}^{N}\pi_i\lambda^{\textrm{max}}(P_i^\alpha\widetilde{E}_i^\alpha P_i^\alpha)=\piv\cdot\mathbf{s}'_\alpha,
\label{eq:ua}
\end{equation}
where, as before, $\piv=\{\pi_1,\pi_2,\dots,\pi_N\}$, and $\mathbf{s}'_\alpha=\{\lambda^{\textrm{max}}(P_1^\alpha\widetilde{E}_1^\alpha P_1^\alpha),\dots,\lambda^{\textrm{max}}(P_N^\alpha\widetilde{E}_N^\alpha P_N^\alpha)\}$ in decreasing order of value (this, actually, defines the labeling of the POVM elements~$\widetilde E_j^\alpha$). Note that this ordering ensures maximization of  the overlap $\piv\cdot\mathbf{s}'_\alpha$.
The probability of success of the optimal message discrimination protocol~is
\begin{equation}
p_s=\max_\alpha p_s^\alpha=\max_\alpha \piv\cdot\mathbf{s}'_\alpha.
\label{ebc31.03.11-2}
\end{equation}
Here $\alpha$ takes $({N+1})^M/N!$ different values, namely, the number of different  ways of distributing $M$ POVM elements in $N+1$ sets, where the sum of the elements in each of these sets are $\widetilde E_1^\alpha$, \dots, $\widetilde E_N^\alpha$,~$\widetilde E_?^\alpha$ respectively  ($N!$~takes into account the specific labeling defined above). Note that
certain sets may be empty, i.e., we allow some of the new POVM
elements to be the zero operator (the corresponding message will never be identified in these cases).

To compute $p_s$ we may consider the following procedure. Pick a grouping $\alpha$ and construct each of the projectors~$P_i^\alpha$ on the
intersection ${\mathscr K}_i^\alpha$ for $i=1,2,\dots,N$. If some  ${\mathscr K}_i^\alpha$ is empty, terminate the
calculation and consider a different grouping $\alpha'$. If there is an empty intersection for all $\alpha$,
the problem does not have a solution (other than the trivial $\widetilde E_?=\id$), which means that the given POVM $\{E_i\}$ cannot be used to unambiguously discriminate $N$ messages.
For each grouping such that ${\mathscr K}_i^\alpha\not=\emptyset$, $i=1,2,\dots, N$, compute $\mathbf{s}_\alpha$ and pick up the one, $\alpha^*$, that maximizes~(\ref{ebc31.03.11-2}).
%
%
Optimal detection is
attained with the POVM measurement $\{ \widetilde{E}_1^{\alpha^*},...,\widetilde{E}_N^{\alpha^*},\widetilde{E}_?^{\alpha^*}\}$ and the
optimal encoding of each classical
messages $i$ is provided by an eigenstate $\rho_i$  of $P_i^{\alpha^*}\widetilde{E}_i^{\alpha^*}P_i^{\alpha^*}$ with
maximal eigenvalue (note that the states $\rho_i$
can always be chosen to be pure).

The above solution to the reverse unambiguous
discrimination problem works for any POVM. In contrast, there is no known
solution to the straight version of the same problem for an arbitrary
ensemble of mixed input states (see, e.g., Ref.~\cite{Raynal05}). As
in the case of minimum error discrimination, there are certain
similarities between the problem of finding the optimal encoding for
a given POVM and that of finding the optimal POVM for a given encoding: for the latter, the POVM $\{\widetilde{E}_i\}$ have to be
chosen such that $\widetilde{E}_i\in \cap_{j\not=i}^N \ker \rho_j$, which resembles the condition $\rho_i\in\cap_{j\not=i}^N\ker \widetilde E_j$ in the reverse problem. Furthermore, in the two problems, one has to maximize the same quantity, Eq.~\eqref{P}, where states and POVM elements play essentially the same role (they are interchangeable). Recall, however, that in the straight case
optimization has the
additional constraint $\sum_i^N\widetilde{E}_i\leq I$, which makes the
problem more difficult.






\section{Mutual information \label{secmut}}

The problems considered in the previous sections characterize the ability of a POVM measurement to perform certain information readout tasks (e.g., minimum error discrimination or unambiguous message discrimination) with respect to a given source of classical messages described by the prior probabilities $\{\pi_i\}$. These results are strongly dependent on the source. For example, if the source consists of only a single message, each of the tasks can be accomplished with unit probability using any measurement. Such a source, however, is trivial as it contains no information. In this section, we consider a source-independent characterization of the ability of a measurement to extract information which is provided by the maximum mutual information that can be established between the sender and the receiver over all possible sources and suitable encodings at the sender's side for the given POVM measurement at the receiver's side.

Consider an information source characterized by the probability distribution $\{\pi_i\}$, $i=1,...,N$, and an encoding $i\rightarrow \rho_i$. The joint probabilities of the input messages and the outcomes of the POVM measurement~$\{E_j\}$, $j=1,...,M$, are
\begin{equation}
P_{ij}=\pi_i\tr(\rho_iE_j).\label{jointprob}
\end{equation}
The mutual information between the input and the output is given by
\begin{equation}
I(P)=\sum_i\eta\!\!\left(\sum_jP_{ij}\right)+\sum_j\eta\!\!\left(\sum_iP_{ij}\right)
-\sum_{ij}\eta(P_{ij}),\label{mutualinfo}
\end{equation}
where $\eta(x)=-x\log x$.

We will be interested in the maximum of $I(P)$ over all possible source distributions $\{\pi_i\}$ and encoding strategies~$i\rightarrow \rho_i$, that is, over all input ensembles $\{\pi_i,\rho_i\}$,
\begin{equation}
C(\{E_i\})=\max_{\{\pi_i,\rho_i\}}I(P).\label{C}
\end{equation}

Note that, according to the data processing inequality, post-processing of information at the receiver's side cannot increase the mutual information, so in this case it cannot help to group POVM elements (or randomize outcomes).

As shown by the following proposition, $C(\{E_i\})$ has a natural interpretation as the \textit{capacity} of the measurement $\{E_i\}$ which for all practical purposes can be modeled by a quantum channel of the form $\mathcal{E}(\rho)=\sum_j\tr(\rho E_j)|j\rangle\langle j|$, where $\{|j\rangle\}$ are orthogonal states that carry the classical information about the outcome of the measurement.

\textit{Proposition 1.} $C(\{E_i\})$ is equal to the classical capacity of the channel
\begin{equation}
\mathcal{E}(\rho)=\sum_j\tr(\rho E_j)|j\rangle\langle j|.
\end{equation}

\textit{Proof.} It is known \cite{Holevo96,SW97} that the classical capacity of a quantum channel $\mathcal{M}$ over independent uses of the channel (i.e., when no entanglement between multiple inputs to the channel is allowed) is given by the quantity
\begin{equation}
\chi(\mathcal{M})\!=\!\!\!\max_{\{\pi_i,\rho_i\}}\!\!\left\{\!S\!\!\left[\!\sum_i\!\pi_i\mathcal{M}(\rho_i)\!\right]\!\!-\!\!\sum_i\pi_iS\!\left[\mathcal{M}(\rho_i)\right]\!\right\},\label{chi}
\end{equation}
where $S(\rho)=-\tr( \rho\log\rho)$ denotes the von Neumann entropy of the state $\rho$. The general capacity of the channel, allowing possibly entangled inputs, is
\begin{equation}
C(\mathcal{M})=\lim_{n\rightarrow\infty}\frac{\chi(\mathcal{M}^{\otimes n})}{n},
\end{equation}
where $\mathcal{M}^{\otimes n}$ denotes $n$ uses of the channel. For entanglement-breaking channels \cite{HSR03}, such as the quantum-to-classical channel $\mathcal{E}(\rho)$ above, it has been shown that the quantity $\chi(\mathcal{E})$ is additive \cite{Shor02,Holevo98,King01}, in particular $\chi(\mathcal{E}\otimes \mathcal{E})=2\chi(\mathcal{E})$, which implies that
\begin{equation}
C(\mathcal{E})=\chi(\mathcal{E}).
\end{equation}
Furthermore, for any input ensemble $\{\pi_i,\rho_i\}$, the channel~$\mathcal{E}(\rho)$ outputs an ensemble of commuting quantum states, $\{\pi_i,\mathcal{E}(\rho_i)\}$, and for such an ensemble it is easy to verify that the quantity $S\left[\sum_i\pi_i\mathcal{E}(\rho_i)\right]-\sum_i\pi_iS[\mathcal{E}(\rho_i)]$ is equal to the mutual information in Eq.~\eqref{mutualinfo}. The proposition then follows from the definitions \eqref{C} and \eqref{chi}.

A comment is in order here. The classical capacity of a channel is the maximum rate at which information can be transmitted reliably through the channel in the limit of infinitely many uses. Since the optimal measurement for extracting information from the channel $\mathcal{E}(\rho)$ is a projective measurement in the basis $\{|j\rangle\}$, which preceded by $\mathcal{E}(\rho)$ is equivalent to the POVM measurement $\{E_j\}$, the quantity $C(\{E_j\})$ is equal to the maximum rate at which information can be read reliably using the POVM~$\{E_j\}$.

\textit{Corollary 1.} We have,
\begin{equation}
C(\{E_i\})\!=\!\!\!\max_{\{\pi_i,\rho_i\}}\!\!\left\{\!S\!\!\left[\!\sum_i\!\pi_i\mathcal{E}(\rho_i)\!\right]\!-\!\sum_i\!\pi_iS[\mathcal{E}(\rho_i)]\right\}.
\end{equation}

Observe that we can write the joint probability \eqref{jointprob} in the symmetric form
\begin{equation}
P_{ij}=\tr(\breve{\rho}_i{E}_j),
\end{equation}
where $\breve{\rho}_i\equiv\pi_i\rho_i$ are \textit{unnormalized} positive operators satisfying $\tr(\sum_i\breve{\rho}_i)=1$. (Hereafter, we will use the notations $\{\pi_i,\rho_i\}$ and $\{\breve{\rho}_i\}$ interchangeably to denote an ensemble of states.) In this notation, $C(\{E_i\})=\max_{\{\breve{\rho}_i\}}I(P)$. Notice further that the mutual information $I(P)$ is symmetric with respect to the indexes $i$ and~$j$. Therefore, the problem we are considering can be regarded as dual to the one of accessible information of an ensemble of states $\{{\breve{\rho}}_i\}$ \cite{Fuchs96} which can be written as
\begin{equation}
A(\{\breve{\rho}_i\})=\max_{\{{E}_i\}}I(P).
\end{equation}
Note, however, that the two problems are not identical as the operators $\{{E}_i\}$ satisfy a stronger constraint than the operators $\{\breve{\rho}_i\}$: $\sum_i{E}_i={I}$. (A strict duality transformation between signal ensembles and POVM measurements has been established in Refs.~\cite{HJW,Hall}. We will not be concerned with that correspondence here.)

The above suggests that certain results in the study of the accessible information of an ensemble of states may prove useful for the study of the capacity of a measurement. For example, the symmetry of the problems and the difference in constraints implies
\begin{equation}
C(\{E_i\})\geq A(\{\breve{E}_i\}),
\end{equation}
where $\breve{E}_i=E_i/d$. Therefore, any known lower bound of $A$ can be used to obtain a lower bound of $C$. For example, the lower bound obtained in Ref.~\cite{JRW94} yields
\begin{equation}
C(\{E_i\})\geq Q\left(\sum_im_i\bar{E}_i\right)-\sum_im_iQ(\bar{E}_i),
\end{equation}
where $m_i=\tr(E_i)/d$, $\bar{E}_i=E_i/(m_id)$, and $Q(\rho)$ is the \textit{subentropy} of a density matrix $\rho$, which in terms of the eigenvalues $\lambda_k$ of $\rho$ reads \cite{JRW94}
\begin{equation}
Q(\rho)=-\sum_k\left(\prod_{l\neq k}\frac{\lambda_k}{\lambda_k-\lambda_l} \right)\lambda_k\log\lambda_k
\end{equation}
(if two or more eigenvalues are equal, one takes the limit as they become equal).

Similarly, one may wonder if the Holevo quantity $S(\sum_im_i\bar{E}_i)-\sum_im_iS(\bar{E}_i)$ \cite{Kholevo79}, which provides a simple upper bound on the accessible information $A(\{\breve{E}_i\})$, could also provide a useful bound for the capacity $C(\{E_i\})$. As we will see below, however, this quantity is neither an upper nor a lower bound to $C(\{E_i\})$.

\textit{Proposition 2.} The maximum in Eq.~\eqref{C} can be achieved
with an ensemble of pure input states
$\rho_i=|\psi_i\rangle\langle\psi_i|$. Furthermore, the number $N$
of input states can be made to satisfy $d\leq N\leq d^2$.

This proposition is similar to Theorem 3 in
Ref.~\cite{Davies78}, where it is shown that for a given ensemble of
input states, the optimal POVM measurement can be taken to have
rank-one POVM elements whose number $M$ satisfies $d\leq M\leq d^2$.

\textit{Proof.} As noted in Ref.~\cite{Davies78}, $I(P)$ is a convex
function over the convex set of $N\times M$ probability matrices $P$
with fixed row sums. By a similar argument, $I(P)$ is a convex
function over the convex set of $N\times M$ probability matrices $P$
with fixed column sums.  This implies that if $P'$ is a $(N-1)\times
M$ probability matrix obtained from $P$ by replacing two rows by
their row sum, then $I(P')\leq I(P)$, with equality when the two
rows are proportional. Therefore, for any input ensemble
$\{\pi_i,\rho_i\}$, where
$\rho_i=\sum_{k}p_{ik}|\psi_{ik}\rangle\langle\psi_{ik}|$, we can
consider the pure-state ensemble
$\{\pi_ip_{ik},|\psi_{ik}\rangle\langle\psi_{ik}|\}$ which has
mutual information with the output no less than that of
$\{\pi_i,\rho_i\}$. (Note that we can assume that no two states
$|\psi_{ik}\rangle\langle \psi_{ik}|$ are identical, since if they
are, we can combine them into a single state with prior probability
equal to the sum of their prior probabilities, which does not change
the mutual information.) Hence, the maximum in Eq.~\eqref{C} is
attained for an ensemble of different pure states.

Next, observe that Eq.~\eqref{C} can be written as
\begin{equation}
C(\{E_i\})=\max_{\rho}\max_{\{\pi_i,\psi_i\}_{\rho}}I(P),\label{C2}
\end{equation}
where the left maximization is over all density matrices $\rho$, and
the right maximization is over all ensembles
$\{\pi_i,\psi_i\}_{\rho}$ of pure states $\psi_i\equiv
|\psi_i\rangle\langle\psi_i|$, whose averages are equal to $\rho$,
$\sum_i\pi_i|\psi_i\rangle\langle\psi_i|=\rho$. (We note that the quantity $\max_{\{\pi_i,\psi_i\}_{\rho}}I(P)$ for a fixed $\rho$ has been previously considered in relation to methods for obtaining bounds on the mutual information \cite{Hall}.) Following closely the proof in Ref.~\cite{Davies78}, we will show that for any $\rho$, $\max_{\{\pi_i,\psi_i\}_{\rho}}I(P)$ can be
achieved by an ensemble of at most $d^2$ states. Indeed, the latter maximization is equivalent to a maximization over the convex set $Y$ of probability distributions with finite support on the set of pure states, whose average is equal to $\rho$. Note that the different ensembles $\{\pi_i,\psi_i\}_{\rho}$ give rise to joint probability matrices $P$ with fixed row sums equal to
$\tr(\rho E_j)$, which according to the convexity property pointed out earlier implies that $I(P)$ is a convex function on $Y$. Hence, the maximum is achieved for an extreme point of $Y$, which by Caratheodory's theorem can be shown to be a probability distribution whose support has $\leq 1+\dim{\mathcal{A}}$ points, where $\mathcal{A}$ is the convex set of density operators of which the pure states we are considering are extreme points. Since $\dim{\mathcal{A}}=d^2-1$, we obtain $N\leq d^2$.

To show that in general $d\leq N$, we will use the fact that for every $d$, there are certain types of POVMs for which the optimal $\rho$ in Eq.~\eqref{C2} is full-rank (in particular, we will show below (Theorem 1) that when the POVM is covariant under the irreducible representation of a finite group, the maximum in Eq.~\eqref{C2} is achieved for $\rho=I/d$). If we assume that $d>N$, there must exist a vector $|\psi\rangle$, $\langle\psi|\psi\rangle=1$, such that $\langle\psi|\psi_i\rangle=0$, $\forall i=1,...,N$. But then $\langle \psi|\rho|\psi\rangle=\sum_i\pi_i|\langle\psi|\psi_i\rangle|^2=0$, which is in contradiction with $\rho$ being full-rank. \hfill$\Box$

We next consider the case of a group covariant POVM measurement, which is dual to the problem of accessible information for a group covariant input ensemble \cite{Davies78}. For this purpose, we need to introduce some terminology. Let $S$ denote the set of all states on a Hilbert space $\mathcal{H}$ of dimension $d$. Following Ref.~\cite{Davies78}, we will regard a representation $R$ of a group $G$ as a homomorphism from $G$ to the affine automorphisms of $S$, where every such automorphism is representable in the form $\alpha(\rho)=U\rho U^{\dagger}$ with $U$ being a unitary or an antiunitary operator (we will consider the action of $R$ automatically extended to all operators over $\mathcal{H}$ by linearity). A representation of $G$ is irreducible, if the only $G$-invariant point of $S$ is $I/d$.

We will say that the POVM $\{E_j\}$, $j=1,...,M$, is $G$-covariant if there exists a surjection $f: G \rightarrow \{E_j\}$, where we denote $f(g):=E_g$, such that $R_g(E_h)=E_{gh}$, $\forall g,h\in G$. Note that every element $E_j$ must equal $E_g$ for at least one $g\in G$, but this correspondence may be degenerate, i.e., a given $E_j$ may be associated with two or more elements of the group. The fact the $G$ is a group implies that this degeneracy must be the same for every element $E_j$, and hence $M$ must be a factor of $|G|$.

\textit{Theorem 1 (The group covariant case).} If the POVM $\{E_j\}$ is covariant with respect to the finite group $G$ that has an irreducible representation $R$ on $S$, then there exists a pure state $|\psi\rangle\langle\psi|$, $\langle\psi|\psi\rangle=1$, such that the maximum in Eq.~\eqref{C} is achieved by the covariant ensemble of pure input states $\{|G|^{-1},R^*_g(|\psi\rangle\langle\psi|)\}$, where $|G|$ is the number of elements of $G$, and $R^*$ denotes the representation of $G$ dual to $R$. The capacity of $\{E_j\}$ is
\begin{equation}
C(\!\{E_j\}\!)\!=\!\log d+M^{-1}\!d\sum_j\langle\psi|\frac{E_j}{\tr E_j}|\psi \rangle\log {\langle\psi|\frac{E_j}{\tr E_j}|\psi \rangle}.\label{Ccov}
\end{equation}

\textit{Proof.} Let $\{\pi_i,\psi_i\}$ be an ensemble of pure input states that maximizes the mutual information for the given covariant POVM measurement $\{E_j\}$. Construct a new input ensemble $\{\tilde{\pi}_{ig},\tilde{\psi}_{ig}\}$, where
\begin{equation}
\tilde{\psi}_{ig}=R^*_g(\psi_i), \mbox{ and }
\tilde{\pi}_{ig}={\pi_i}|G|^{-1}.
\end{equation}
The new probability matrix $\tilde{P}$ obtained using this ensemble has the form
\begin{equation}
\tilde{P}=|G|^{-1}\left(\begin{array}{l}
P_1\\
P_2\\
\vdots\\
P_{|G|}
\end{array}\right),
\end{equation}
where each of the probability matrices $P_1, P_2,...,P_{|G|}$ is obtained from $P$ by a permutation of the rows and columns of $P$, and the column sums of $\tilde{P}$ are all equal to~$|G|^{-1}$. A straightforward calculation shows that the new probability matrix yields a value for the mutual information which is no less than that obtained for $P$, i.e.,~$I(\tilde{P})\geq I(P)$:
\begin{eqnarray}
I(\tilde{P})\!\!&\equiv&\!\!\sum_i\eta\!\!\left(\!\sum_j\tilde{P}_{ij}\!\!\right)\!+\!\sum_j\eta\!\left(\!\sum_i\tilde{P}_{ij}\!\!\right)
\!-\!\sum_{ij}\eta(\tilde{P}_{ij})\nonumber\\
&=&\!\! |G|\sum_i\eta\!\!\left(\sum_j|G|^{-1}{P}_{ij}\!\!\right)\!+\!\sum_j\eta\left(|G|^{-1}\right) \nonumber\\
&-&\!\!|G|\sum_{ij}\eta(|G|^{-1}{P}_{ij}) \nonumber\\
&=&\!\!\left[ \sum_i\eta\!\!\left(\sum_j{P}_{ij}\right)\!+\log|G|^{-1}\right]\!+\!\sum_j\eta\!\left(|G|^{-1}\right) \nonumber\\
&-&\!\!\left[\sum_{ij}\eta({P}_{ij})+\log|G|^{-1}\right] \nonumber\\
&\geq&\!\! \sum_i\eta\!\!\left(\sum_j{P}_{ij}\right)\!+\sum_j\eta\!\left(\sum_i{P}_{ij}\!\right) \nonumber\\
&-&\!\!\sum_{ij}\eta({P}_{ij})\equiv I(P).
\end{eqnarray}
Now, consider the covariant input ensembles $\{|G|^{-1}, \bar{\psi}^i_g \}$, where
\begin{equation}
\bar{\psi}^i_g=R^*_g(\psi_i).
\end{equation}
Let us denote the probability matrices that each of these ensembles yields by $\bar{P}_i$. Since the ensemble $\{\tilde{\pi}_{ig},\tilde{\psi}_{ig}\}$ is a convex combination of the ensembles $\{|G|^{-1}, \bar{\psi}^i_g \}$, and the mutual information is a convex function of the input ensemble, we obtain
\begin{equation}
I(P)\leq I(\tilde{P})\leq \max_i{I(\bar{P}_i)},
\end{equation}
i.e., the maximum in Eq.~\eqref{C} is achieved for one of the covariant input ensembles $\{|G|^{-1}, \bar{\psi}^i_g \}$ which has the form stated in the theorem. The value of the capacity [Eq.~\eqref{Ccov}] is obtained by a straightforward calculation, taking into account the possible degeneracy in the correspondence between the group elements and the POVM elements.  \hfill$\Box$

Notice that since the average of a group covariant ensemble is $G$-invariant, from the irreducibility of $R$ it follows that $\sum_g|G|^{-1}\bar{\psi}^i_g=I/d$. This shows that indeed for every $d$ there are POVM measurements that require at least $d$ optimal input states as argued in the proof of Proposition 2.

\textit{Comment.} The optimal ``seed'' $|\psi\rangle\langle\psi|$ may be such that the input ensemble $\{|G|^{-1},R^*_g(|\psi\rangle\langle\psi|)\}$ contains identical states, i.e., it may be that $R^*_g(|\psi\rangle\langle\psi|)=R^*_h(|\psi\rangle\langle\psi|)$ for certain $g\neq h$. The fact that $G$ is a group implies that each maximal set of identical states in the ensemble must contain the same number of elements (and hence the number $N$ of distinct states in the ensemble must be a factor of $|G|$). It is straightforward to see that the ensemble $\{N^{-1}, |\psi_i\rangle\langle\psi_i|\}$ obtained from $\{|G|^{-1},R^*_g(|\psi\rangle\langle\psi|)\}$ by identifying the identical states and redefining their probabilities as the sum of the original probabilities, is also optimal. This is because the joint probabilities resulting from the input ensemble $\{N^{-1}, |\psi_i\rangle\langle\psi_i|\}$ can be transformed into those resulting from $\{|G|^{-1},R^*_g(|\psi\rangle\langle\psi|)\}$ by local postprocessing on the sender's side, which cannot increase the mutual information. Hence, the number of states in the optimal ensemble in general may be smaller than $|G|$ (just as the number of outcomes of a group covariant POVM may be smaller than $|G|$). This is the case, for example, with the optimal ensemble for the two-dimensional SIC-POVM studied in Section \ref{sec:mutual}, which has 4 elements while the symmetry group has 12.

\textit{Corollary 2.} In the group covariant case, we have
\begin{equation}
C(\{E_i\})=A(\{\breve{E}_i\}).
\end{equation}
Moreover, if the POVM measurement $\{F_j\}$ optimizes the mutual information for the input ensemble $\{\breve{E}_i\}$, the input ensemble $\{{\breve{F}_j}\}$, where $\breve{F}_j\equiv F_j/{d}$, optimizes the mutual information for the measurement $\{E_i\}$.

Since under this symmetry the problem is equivalent to that of accessible information of a covariant input ensemble, any known results in the latter case can be applied here (see, e.g., Ref.~\cite{Davies78}).
In particular, in Section \ref{sec:mutual} we calculate the capacity of the two-dimensional SIC-POVM.

Another important case in which calculating the capacity of a measurement reduces to a well known problem is that of a POVM $\{E_i\}$ with commuting elements,~$[E_i,E_j]=0$, $\forall i,j$. In this case, we can assume that the optimal signal states $\rho_i$ are diagonal in the eigenbasis of $\{E_j\}$, since for any $\rho$, the state $\rho'=\textrm{diag}({\rho}_{nn})$, where $\rho_{nn}$ are the diagonal elements of $\rho$ in the eigenbasis of $\{E_j\}$, yields the same values for the joint probabilities $\tr(\rho E_j)$. Furthermore, as we saw in the proof of Proposition 2, the optimal input ensemble can be taken to consist of the eigenstates of all $\rho_i$, which means that the maximum in Eq.~\eqref{C} is achieved for an ensemble of input states which are the common eigenbasis of $\{E_i\}$. Hence the joint probabilities are $P_{ij}=\pi_i\lambda^i_j$, where $\lambda^i_j$ is the $i$-th eigenvalue of $E_j$, and the problem reduces to finding $\max_{\{\pi_i\}}I(P)$ which is the capacity of the classical channel described by the conditional probability matrix~$p(j|i)=\lambda^i_j$. Note that a measurement with two outcomes necessarily has commuting POVM elements, i.e., the capacity of a two-outcome measurement is always equal to the capacity of a classical channel with a binary output. Thus, for example, the capacity of a two-outcome qubit measurement that has elements~$E_1=\textrm{diag}(\alpha,\beta)$, $E_2=\textrm{diag}(1-\alpha,1-\beta)$ in some basis can be obtained from the formula for the capacity of a general binary channel \cite{Reza61}
\begin{eqnarray}
C(\alpha,\!\beta)&=& \frac{\alpha H(\beta)-\beta H(\alpha)}{\beta -\alpha}
\nonumber\\
&+&\log\left[1\!+\!\exp\frac{H(\alpha)\!-\!H(\beta)}{\beta-\alpha}\right],
\end{eqnarray}
where $H(q)=-q\log{q}-(1-q)\log(1-q)$, $q\in[0,1]$, is the entropy of a binary source.
The optimal prior distribution in this case is $\{p,1-p\}$, where \cite{Reza61}
\begin{equation}
p=\frac{\beta}{\beta-\alpha}-\frac{1}{(\beta-\alpha)\left[1+\exp\frac{H(\beta)-H(\alpha)}{\beta-\alpha}\right]}.\label{optinbin}
\end{equation}

We can now see that, as mentioned earlier, the naively constructed Holevo quantity $S(\sum_im_i\bar{E}_i)-\sum_im_iS(\bar{E}_i)$ where $m_i=\tr(E_i)/d$, $\bar{E}_i=E_i/(m_id)$, in general is neither an upper nor a lower bound to $C(\{E_i\})$. Indeed, it is known that the accessible information of an ensemble of density matrices is equal to the Holevo quantity of the ensemble if and only if all density matrices in the ensemble commute, and the maximal value of the mutual information is attained for a projective measurement in the common eigenbasis of the input ensemble. From the symmetry of the problem we see that for a POVM with commuting elements, the quantity~$S(\sum_im_i\bar{E}_i)-\sum_im_iS(\bar{E}_i)$ is equal to the mutual information between the equiprobable input ensemble of common eigenstates of $\{E_i\}$ and the outputs of the measurement $\{E_i\}$. However, from Eq.~\eqref{optinbin} it can be seen that an equiprobable prior distribution is generally suboptimal for this case, i.e., the quantity~$S(\sum_im_i\bar{E}_i)-\sum_im_iS(\bar{E}_i)$ can be strictly smaller than $C(\{E_i\})$. On the other hand, in the group covariant case we have~$C(\{E_i\})=A(\{\breve{E}_i\})$, where in general $A(\{\breve{E}_i\})$ is strictly smaller than $S(\sum_im_i\bar{E}_i)-\sum_im_iS(\bar{E}_i)$.

We remark that the maximal possible mutual information for an input ensemble of states on a Hilbert space of dimension $d$ and any POVM measurement is $\log d$. This can be easily seen from Holevo's upper bound on the accessible information \cite{Kholevo79}. Moreover, this quantity is achievable only by an ensemble of pure commuting input states that sum up to the maximally mixed state, i.e., by an equiprobable ensemble of orthogonal basis states. The unique optimal measurement for such an ensemble is a projective measurement on the basis in question. Reversely, any rank-one projective measurement has capacity $\log d$ which is achievable by the equiprobable input ensemble of corresponding basis states. Hence, rank-one projective measurements have the highest capacity.

\section{Example: The SIC-POVM on a qubit}

In this section, we apply the above results to the case of a symmetric informationally complete (SIC) POVM on a qubit, as well as to a noisy, or unsharp, version of this POVM. A SIC-POVM \cite{Zauner99,Renes04} in dimension $d$ consists of a set of $d^2$ rank-one positive operators, $E_i=(1/d)|\psi_i\rangle\langle\psi_i|$, where  the pure states $|\psi_i\rangle$ are such that~$|\langle\psi_i|\psi_j\rangle|^2=1/(d+1)$ for $i\not=j$. The measurement is called ``complete'' in the sense that its statistics is sufficient for the full tomography of any quantum state~\cite{Prugovecki77,Busch91}. SIC-POVMs are of particular interest due to their various applications in quantum information, including quantum tomography~\cite{Caves02}, quantum cryptography \cite{Fuchs03}, and the foundations of quantum mechanics \cite{Fuchs02}.

Up to a change of basis, the POVM elements of such a measurement for $d=2$ can be written as
\begin{equation}
E_i=\frac{1}{4}(I+\vec{n}_i\cdot\vec{\sigma}),\hspace{0.2cm}i=1,2,3,4,
\label{eq:SIC-POVM}
\end{equation}
where $\vec{\sigma}$ is the vector of Pauli matrices
\begin{equation}
\sigma_y=\left(\begin{array}{cc}
0&1\\
1&0
\end{array}\right), \hspace{0.1cm}\sigma_y=\left(\begin{array}{cc}
0&-i\\
i&0
\end{array}\right),\hspace{0.1cm}\sigma_z=\left(\begin{array}{cc}
1&0\\
0&-1
\end{array}\right),
\end{equation}
and
\begin{eqnarray}
&\overrightarrow{n}_1=\frac{1}{\sqrt{3}}\left(\begin{array}{c}
1\\1\\1
\end{array}\right), \hspace{0.2cm}\vec{n}_2=\frac{1}{\sqrt{3}}\left(\begin{array}{c}
-1\\-1\\1
\end{array}\right),\nonumber\\
&\vec{n}_3=\frac{1}{\sqrt{3}}\left(\begin{array}{c}
-1\\1\\-1
\end{array}\right),\hspace{0.2cm}\vec{n}_4=\frac{1}{\sqrt{3}}\left(\begin{array}{c}
1\\-1\\-1
\end{array}\right).
	\label{eq:bloch}
\end{eqnarray}

In order to illustrate the relation between the ``sharpness'' of a measurement and its ability to read out information, we will consider a more general, noisy version of the above SIC-POVM, where each outcome is mixed with some amount of white noise,
\begin{eqnarray}
{E}_i(\epsilon)=\epsilon E_i+(1-\epsilon) \frac{I}{4}=\frac{1}{4}(I+\epsilon\vec{n}_i\cdot\vec{\sigma}), \hspace{0.3cm} \label{fuzzy}\\
i=1,2,3,4, \hspace{0.3cm}0\leq \epsilon \leq 1.\nonumber
\end{eqnarray}
When $\epsilon=1$, the measurement reduces to the ideal SIC-POVM [$E_i(1)\equiv E_i$], while as $\epsilon\rightarrow 0$, the measurement becomes infinitesimally weak \cite{weak}, approaching a trivial measurement, each of its outcomes occurring with probability $1/4$ independently of the input state. In this sense, $\epsilon$ can be regarded as parameterizing the ``sharpness'' or ``strength'' of the measurement \cite{rapcan11}.

\subsection{Minimum error discrimination}

For simplicity, let us start with the noiseless SIC-POVM  (\ref{eq:SIC-POVM}).
Given the symmetry of the problem, it is enough to consider four groupings, $\alpha\in\{A,B,C,D\}$:
\begin{eqnarray}
&A: \{E_{1},E_{2},E_{3},E_{4}\} \nonumber \\
&B: \{E_{1}+E_{2},E_{3},E_{4},0\}    \nonumber\\
&C: \{E_{1}+E_{2},E_{3}+E_{4},0,0\}   \nonumber\\
&D: \{E_{1}+E_{2}+E_{3},E_{4},0,0\} .
\label{ebcSUMS}
\end{eqnarray}
The corresponding vector of maximum eigenvalues (in decreasing order of value) are
[see Eq.~(\ref{Blambdamax}) with $B_{ij}=\delta_{ij}$]
\begin{eqnarray}
& \mathbf{s}_{A}=\{1/2,1/2,1/2,1/2\} ,\nonumber \\
& \mathbf{s}_{B}=\{(1+1/\sqrt{3})/2,1/2,1/2,0,0\} ,\nonumber\\
& \mathbf{s}_{C}=\{(1+1/\sqrt{3})/2,(1+1/\sqrt{3})/2,0,0\} ,\nonumber\\
& \mathbf{s}_{D}=\{1,1/2,0,0\},
\label{eq:grouping}
\end{eqnarray}
where it is understood that the vectors need to be padded with extra zeros if the number of signal states exceeds four ($N>4$).
For equiprobable signals, $\pi_{i}=1/N$, the optimal success probability is given by $p_{s}=1/N \max_{\alpha}\sum_{i=1}^{N} (\mathbf{s}_{\alpha})_{i}$. In particular,  $p_{s}=1/2+1/(2\sqrt{3})$ for $N=2$, $p_{s}=1/2+1/(6\sqrt{3})$ for $N=3$, and $p_{s}=2/N$ for $N\geq 4$, which are attained by the groupings $C$, $B$ and $A$, respectively. That is, for four signals ($N=4$) no grouping is necessary and the signal states have to be chosen to point along the directions of the SIC-POVM \eqref{eq:bloch}. Any additional signals ($N>4$) can be assigned to arbitrary states and will never contribute to the success probability. For $N=3$ one has to group two POVM elements leading to un unsharp effective measurement, and leave the remaining two outcomes ungrouped (i.e., sharp). In that case the three signals lie on a plane: two signals point along, say,  $\vec{n}_1$ and $\vec{n}_2$ (corresponding to the sharp POVM elements), and the third points along $-(\vec{n}_1+\vec{n}_2)$.
For $N=2$ the optimal strategy is two encode the signals into orthogonal states pointing along the directions  resulting from pairwise groupings, e.g., $\vec{n}_1+\vec{n}_2$ and $\vec{n}_3+\vec{n}_4=-(\vec{n}_1+\vec{n}_2)$.

\begin{figure}[htbp] 
   \centering
  \includegraphics[width=3in]{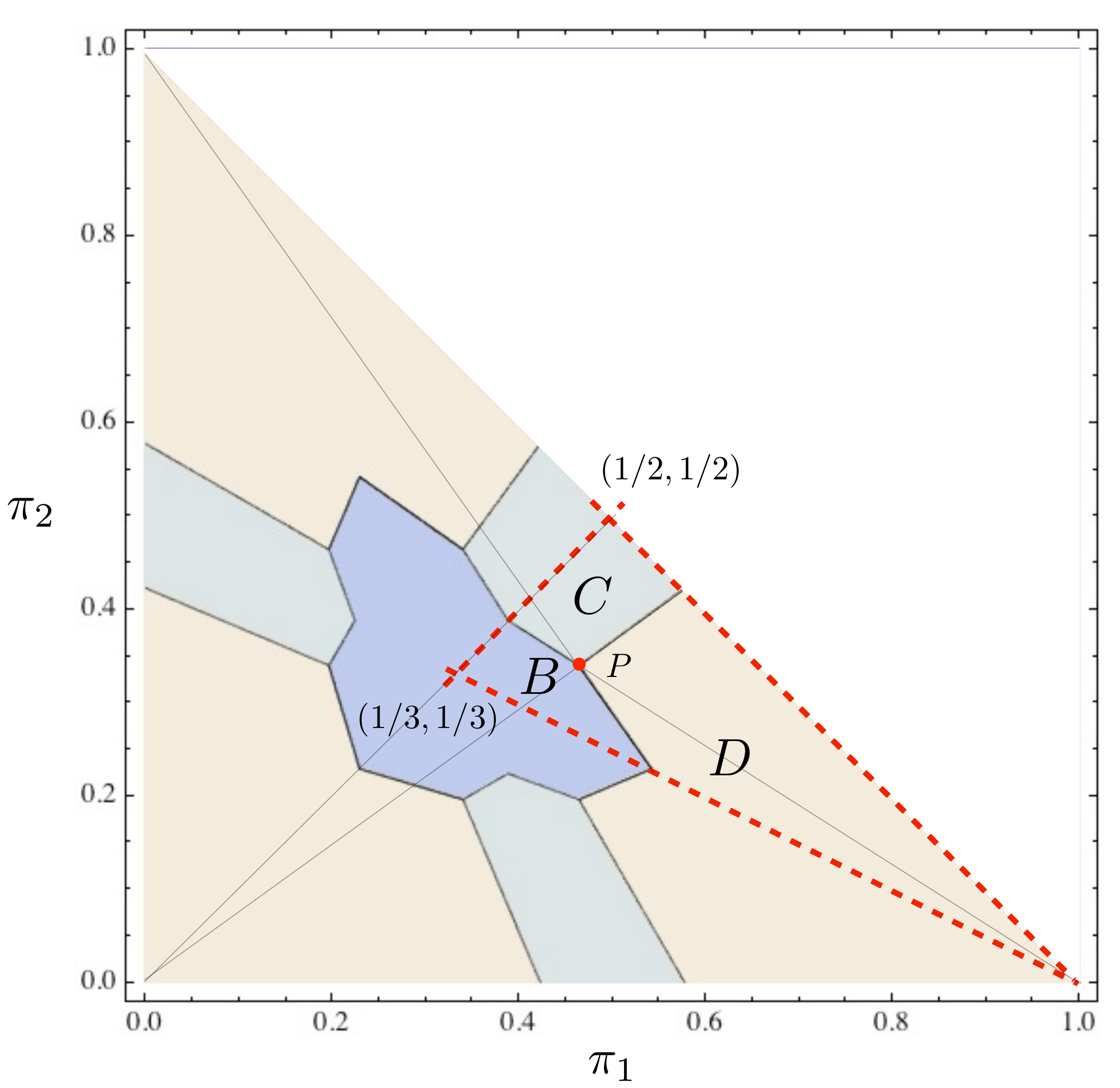}
   \caption{(color online) The colored regions in the prior-probability simplex for $N=3$ indicate the various optimal groupings. The dashed outline delimits the region~\mbox{$\pi_1\ge\pi_2\ge\pi_3$.} Within this region, the intersection point of~$A$, $B$ and $C$ is $P=\{2\sqrt{3}-3,9-5\sqrt{3}\}$. Auxiliary thin lines are drawn to help understanding the figure.}
   \label{fig:plotReg}
\end{figure}

In Figure \ref{fig:plotReg} we show the optimality regions for $N=3$ and different priors.
Within the region $\pi_{1}\geq \pi_{2}\geq \pi_{3}$, delimited by a dashed outline in the figure, we observe that~the set of points where each particular grouping is dominant is a convex polytope.
The corresponding maximum success probabilities are:
\begin{eqnarray}
p_{s}^{B}&=&\frac{1}{2}+\frac{1}{2\sqrt{3}} \pi_{1},\nonumber\\
 p_{s}^{C}&=&\left(\frac{1}{2}+\frac{1}{2\sqrt{3}}\right)(\pi_{1}+\pi_{2}), \nonumber\\
 p_{s}^{D}&=&\pi_{1}+\frac{1}{2}\pi_{2}.
\end{eqnarray}
Note that regions $C$ and $D$ correspond to groupings where no outcome is assigned to the third
signal state.
This illustrates the fact that there are cases (regions $C$ and $D$) where it pays not to assign any measurement-outcome to some of the messages ($i=3$ in this example), even though the source emits them with non-zero prior probability. In particular, if the source is strongly biased towards one message ($\pi_{1}\geq 1/\sqrt{3}$, in this example), all but one measurement outcome will be assigned to it~(message~$i=1$).

In order to study the effect of noise, $\epsilon<1$ in  \eqref{fuzzy}, one proceeds along the same lines as above. We first note that since the noise is isotropic, the optimal signal states, i.e., the eigenvectors with maximum eigenvalue of the sums of POVM elements in each grouping, $A$ to $D$, are the same as those for the sharp case, thus independent of the sharpness parameter $\epsilon$. Their corresponding maximum eigenvalues in\eqref{eq:grouping} now have a noisy component that scales with the number~$k_{\alpha i}$ of elements in those sums. More precisely, the vectors of eigenvalues have now components $ \epsilon\, (\mathbf{s}_{\alpha})_{i}+k_{\alpha i}(1-\epsilon)/4$.

For equiprobable signals, $\pi_{i}=1/N$, the optimal groupings  are those that are optimal in the sharp case. Thus, they do not depend on $\epsilon$, only on the number~$N$ of input states. The minimum errors are now:  $p_{s}=1/2+\epsilon/(2\sqrt{3})$ for $N=2$, $p_{s}=1/3+\epsilon (1+1/\sqrt{3})/6$ for $N=3$, and $p_{s}=(1+\epsilon)/N$ for $N\geq 4$.

In more generic cases, when the source emits symbols with arbitrary prior probabilities, the regions of optimality do depend on the noise or sharpness parameter~$\epsilon$. For the case of ternary sources, $N=3$, it is straightforward to show that the overall structure of the optimality regions is that in Figure~\ref{fig:plotReg}, but the point $P({\epsilon})$ where $B$, $C$ and $D$ intersect, moves monotonically away from $P(1)=P=\{2\sqrt{3}-3,9-5\sqrt{3}\}$ (when the POVM is sharp) to $P(0)=\{1/3,1/3\}$ (when it is maximally unsharp).

\subsection{Unambiguous discrimination}

We now turn to unambiguous discrimination with the SIC-POVM on a qubit. Clearly, the slightest amount of noise ($\epsilon< 1$) will ruin any possibility of performing unambiguous discrimination since any signal state can trigger each of the outcomes with a non-zero probability. We thus concentrate on the ideal sharp SIC-POVM. In a two-dimensional Hilbert space one can only hope to unambiguously discriminate two states ($N=2$;  $\pi_1<1$), hence grouping $A$ can be excluded as it has too many outcomes.
Moreover, we need only to consider groupings $B$ and~$D$, since only they have at least one rank-one POVM element and have, therefore, a non empty kernel (${\mathscr K}_j^\alpha\not=\emptyset$).
If grouping $B$ is used, two messages can be unambiguously identified by choosing the signals in the kernels of $E_{3}$ and $E_{4}$ respectively, that is $\rho_{1}=(1-\vec{n}_3\cdot \vec{\sigma})/2$ and $\rho_{2}=(1-\vec{n}_4\cdot\vec{\sigma})/2$, so that outcome 4 can only be triggered by $\rho_{1}$
and outcome 3 by~$\rho_{2}$ (i.e., $\widetilde E_{1}=E_{4}$,  $\widetilde E_{2}=E_{3}$ and  $\widetilde E_{?}=E_{1}+E_{2}$). This leads to a probability of successful identification given by
\begin{eqnarray}
p_{s}^B&=&\pi_{1} \tr(\widetilde E_{1} \rho_{1})+\pi_{2} \tr(\widetilde E_{2} \rho_{2})\nonumber \\[.5em]
&=& {1-\vec{n}_{4}\cdot \vec{n}_{3}\over4}={1\over3},
\end{eqnarray}
which is independent of the prior probabilities $\{\pi_{1},\pi_{2}\}$.

Proceeding along the same lines, one finds that for grouping  $D$ one can only unambiguously identify
the state $\rho_{1}=(1+\vec{n}_4\cdot \vec{\sigma})/2$ with $\widetilde E_{1}=E_{4}$, by excluding
$\rho_{2}=(1-\vec{n}_4\cdot \vec{\sigma})/2$ (i.e., $\rho_2\in \ker \widetilde E_1$), while all other outcomes of the original POVM will be necessarily inconclusive
 ($\widetilde E_{?}=\id-E_{4}$). Obviously, no outcome will be associated to message $i=2$ ($\widetilde E_2=0$). The success probability is
\begin{equation}
p_{s}^D=\pi_{1}\tr(\widetilde E_{1}\rho_{1})={\pi_{1}\over2},
\end{equation}
which beats that of grouping $B$ for $\pi_{1}>2/3$.

\subsection{Mutual information \label{sec:mutual}}

The SIC-POVM on a qubit, including its noisy version, is covariant under the tetrahedral group (indeed, the tips of the Bloch vectors\eqref{eq:bloch} corresponding to the POVM elements define the vertices of a tetrahedron). Therefore, according to Theorem 1 in Section \ref{secmut}, the mutual information for this POVM is maximized by an ensemble of pure input states possessing the same symmetry. Its maximal value, i.e., the capacity of the measurement, is given by Eq.\eqref{Ccov} for a state $\psi$ from the optimal ensemble (all other states in the ensemble are obtained from $\psi$ by applying operators of the symmetry group, i.e., $\psi$ plays the role of a ``seed'' for the ensemble).

\textit{Theorem 2 (Capacity of the noisy two-level SIC-POVM).} For every value of~${\epsilon\in[0,1]}$, the seed $\psi$ that maximizes expression\eqref{Ccov} can be chosen such that its Bloch vector is anti-parallel to the Bloch vector of any one of the four POVM elements\eqref{fuzzy}, i.e., $\vec v=-\vec n_j$. The capacity of the (generally noisy) SIC-POVM is
\begin{equation}
C_{\epsilon}=1+{1-\epsilon\over4}\log{1-\epsilon\over2}+3{1+\epsilon/3\over4}\log{1+\epsilon/3\over2}.\label{theo2}
\end{equation}

This result, which applies to both the straight and reverse formulations of the problem, is interesting on its own right. As far as we are aware, previous results (for $\epsilon=1$) relied on numerical optimization \cite{Davies78}. Here we provide an analytical proof for  $0\leq\epsilon\leq 1$.

\textit{Proof.}
Let us define
$$
h(t)\equiv \eta\left({1+t\over2}\right).
$$

We will first show that the following inequality holds for $-1\le t\le 1$ and $0\le\epsilon\le1$:
\begin{equation}
h(\epsilon t)\ge a(\epsilon)+b(\epsilon) t+c(\epsilon) t^2\equiv \wp_\epsilon(t),
\label{inequality2}
\end{equation}
where
\begin{eqnarray*}
a(\epsilon)&=&{1\over16}\left[  h(-\epsilon)+15h(\epsilon/3)-4\epsilon h'(\epsilon/3)  \right],\\
b(\epsilon)&=&{1\over8}\left[ -3 h(-\epsilon)+3h(\epsilon/3)+4\epsilon h'(\epsilon/3)  \right],\\
c(\epsilon)&=&{3\over16}\left[  3h(-\epsilon)-3h(\epsilon/3)+4\epsilon h'(\epsilon/3)  \right],\\
\end{eqnarray*}
and $h'$ is the derivative of $h$ with respect to its argument.

\bigskip

We start by noticing the following relations:
\begin{equation}
\wp_\epsilon(-1)=h(-\epsilon),\hspace{0.2cm} \wp_\epsilon(1/3)=h(\epsilon/3),\hspace{0.2cm} \wp'_\epsilon(1/3)=\epsilon h'(\epsilon/3),
\label{byconstruction}
\end{equation}
and
\begin{equation}
\gamma(\epsilon)\equiv c(\epsilon)+{\epsilon^2\over4\ln2}\le0,
\label{gamma}
\end{equation}
where the equality is attained only at $\epsilon=0$.
The first three of them are immediate. The last one is not so obvious and can be proved as follows.
The function $\gamma(\epsilon)$ is concave in $[0,1]$ since
\begin{eqnarray}
\gamma''(\epsilon)=-{9\over2(1-\epsilon)(3+\epsilon)^2\ln2}+{1\over2\ln2}\nonumber\\=-{\epsilon(3+5\epsilon+\epsilon^2)\over2(1-\epsilon)(3+\epsilon)^2\ln2}\le0.\nonumber
\end{eqnarray}
Differentiating the expression of $c(\epsilon)$ above we readily obtain
$$
c'(\epsilon)={3\over16}\left[  -3h'(-\epsilon)+3h'(\epsilon/3)+{4\over3}\epsilon h''(\epsilon/3)\right],
$$
which vanishes at $\epsilon=0$. Thus $\gamma'(0)=\gamma''(0)=0$ and $\gamma''(\epsilon)<0$ if~$ \epsilon>0$. Then,~$\gamma(\epsilon)$ must necessarily decrease for $\epsilon>0$, which in turn implies that $\gamma(\epsilon)$ has its unique maximum at~$\epsilon=0$. Since $\gamma(0)=0$, Eq.~(\ref{gamma}) holds in the whole interval~$[0,1]$.

We can now turn to proving~(\ref{inequality2}). We assume that $\epsilon>0$, since $\epsilon=0$ is a trivial case.
%
If $f(t)=h(\epsilon t)-\wp_\epsilon(t)$, then
$$
f''(t)=-2c(\epsilon)-{\epsilon^2\over2(1+\epsilon\, t)\ln2}.
$$
It follows  from this equation that there is only one value of $t$ at which $f''(t)$ vanishes.
But using~(\ref{gamma}), we see that $f''(t)>0$ for $t\ge0$. Therefore, $f''(t)$ can only change sign at some~$t_0<0$. Hence, $f(t)$ is convex in $(t_0,1]$ and concave in $[-1,t_0)$. It can have only one minimum in $(t_0,1]$, and according to the third relation~(\ref{byconstruction}), it must be at~$t=1/3$. Using the second relation~(\ref{byconstruction}), we see that this minimum value is $0$. Thus~$f(t)\ge0$ if $t\in[t_0,1]$. Because of the concavity of $f$ in the other interval, we just need to check the value of $f$ at the end point $t=-1$ [by continuity we must have~$f(t_0)\ge0$]. The first relation~(\ref{byconstruction}) ensures that~$f(t)\ge0$ also in $[-1,t_0]$.

Now, using the inequality~(\ref{inequality2}), one can show that the mutual information for the POVM \eqref{fuzzy},
$$
I=1-{1\over2}\sum_{j=1}^4\eta\left({\langle\phi|E_j(\epsilon)|\psi\rangle\over{\rm tr} E_j(\epsilon)}\right)=
1-{1\over2}\sum_{j=1}^4h\left(\epsilon\, \vec v\cdot \vec n_j\right),
$$
is bounded as
\begin{eqnarray}
I\le1-{1\over2}\sum_{j=1}^4\wp_\epsilon(\vec v\cdot \vec n_j)=1-{1\over2}\left[4 a(\epsilon)+{4\over3}c(\epsilon)\right]\nonumber\\
=1-{h(-\epsilon)+3h(\epsilon/3)\over2} .\nonumber
\end{eqnarray}
This bound is attained with any one of the four choices $\vec v=-\vec n_j$. The value of the capacity\eqref{theo2} is obtained by a straightforward substitution. \hfill$\Box$

Note that in the minimum error scenario, the optimal signal ensemble is such that each state and its corresponding POVM element have maximum overlap  (i.e., they are aligned to each other). In contrast, here we find that it pays to have a signal ensemble where each state would be excluded by one of the POVM outcomes in the absence of noise (i.e., states and POVM elements are anti-aligned to each other). This configuration minimizes the (average) conditional entropy of the output  (the POVM outcomes) given the input  signal ensemble [recall that the mutual information\eqref{Ccov} can be  obtained by subtracting this conditional entropy from the entropy of the output, which is constant here].

As expected, the capacity attains its maximal value $C_1=\log{4/3}$ for $\epsilon=1$ (the ideal SIC-POVM) and monotonically decreases towards $0$ as $\epsilon$ approaches  $0$. Note that, as pointed out in Corollary 2, the capacity of such a group covariant POVM is equal to the accessible information of an equiprobable ensemble of states proportional to the original POVM elements,
\begin{equation}
\rho_i=\frac{I+\epsilon\vec{n}_i\cdot\vec{\sigma}}{2},\hspace{0.2cm}i=1,2,3,4.
\end{equation}
The latter problem, in the case $\epsilon=1$, was studied in Ref.~\cite{Davies78} where it was shown that the accessible information of the corresponding ensemble is $A=\log{4/3}$, which is equal to $C_1$. The capacity of the ideal SIC-POVM has also been previously obtained in Ref.~\cite{Hall} by a different approach.

\section{Conclusion}

In summary, we have studied the problem of optimal signal states for information readout with a given quantum detector. We considered some of the most common information transmission problems---the Bayes cost problem, unambiguous message discrimination, and the maximal mutual information. We provided solutions to the Bayesian and unambiguous discrimination strategies. We also showed that the maximal mutual information is equal to the classical capacity of the measurement and studied its properties in certain special cases. For a group covariant measurement, we obtained that the problem is equivalent to the problem of accessible information of a group covariant ensemble of states. As an example, we applied our results for the different discrimination strategies to the case of a SIC-POVM on a qubit, including a noisy version of that POVM.

An interesting question for future investigation is if and under what conditions the optimal solutions provided here are unique. Another question of significant interest would be to obtain an upper bound on the capacity of a measurement. We provided a lower bound which is obtained from a lower bound on the accessible information, but that lower bound could also be improved. It would also be interesting to investigate the continuity properties of the optimal quantities considered in this paper. For example, if two measurements are close in terms of the distance functions introduced in Ref.~\cite{Oreshkov09}, are their capacities also close?

Finally, we note that the capacity of a POVM provides a very natural and source-independent means to give a quantitative characterization of a generalized quantum measurement. However, it cannot be used as the unique figure of merit against which measurement devices should be benchmarked. Ultimately, the performance of a given measurement apparatus strongly depends on the task it is meant to accomplish.
For instance, a noisy Stern-Gerlach measurement might have a higher capacity than that of an ideal SIC-POVM, however, it would be misleading to claim that such a Stern-Gerlach measurement outperforms the SIC-POVM since the latter can carry out tasks, such as full single-qubit tomography or unambiguous state discrimination, that are impossible to achieve with the former.

\textbf{Note added.} Almost simultaneously with the posting of this paper, two concurrent works appeared---by M. Dall'Arno, G. M. D'Ariano, and M. F. Sacchi (Ref.~\cite{Sacchi}), and by A. S. Holevo (Ref.~\cite{Holevo11})---which also introduce and study the capacity of a POVM measurement.

\section*{Acknowledgments}
We thank Alex Monras for valuable discussions.
This work was supported by the Spanish MICINN through
the Ram\'{o}n y Cajal program (JC), contract FIS2008-01236, and
project QOIT (CONSOLIDER2006-00019), and by the Generalitat de
Catalunya through CIRIT 2009SGR-0985. OO was supported in part by the Interuniversity Attraction Poles program of the Belgian Science Policy Office, under grant IAP P6-10 $\ll$photonics@be$\gg$. EB thanks the HET group at BNL and  Hunter College of the CUNY for their hospitality during the last stages of this work and acknowledges financial support from the Spanish MICINN, reference number PR2010-0367.

\end{document}